\documentclass[preprint,12pt]{elsarticle}

\usepackage{amssymb}
\usepackage{amsthm}

\journal{Applied Mathematical Modelling}

\usepackage{siunitx}
\usepackage{amsmath}
\newcommand{\tsub}{\textsubscript}

\usepackage[italicdiff]{physics}
\usepackage{multirow}
\usepackage{longtable}

\usepackage{comment}
\usepackage{soul}
\usepackage[dvipsnames]{xcolor}

\usepackage[automake,nonumberlist,style=super,nopostdot]{glossaries-extra}
\newglossary*{symb}{List of Symbols}
\newglossary*{greek}{List of Greek Letters}
\newglossary*{sub}{List of Subscripts}
\newglossary*{sup}{List of Superscripts}
\newglossaryentry{symb:a}{type=symb,name={\ensuremath{a}},description={Generic polynomial coefficient}}
\newglossaryentry{symb:C}{type=symb,name={\ensuremath{C}},description={Generic constant term}}
\newglossaryentry{symb:DEff}{type=symb,name={\ensuremath{D_{Eff}}},description={Effective diffusivity}}
\newglossaryentry{symb:E}{type=symb,name={\ensuremath{E}},description={Voltage}}
\newglossaryentry{symb:eact}{type=symb,name={\ensuremath{e_{act}}},description={Fuel cell's activation voltage loss}}
\newglossaryentry{symb:econc}{type=symb,name={\ensuremath{e_{conc}}},description={Fuel cell's concentration voltage loss}}
\newglossaryentry{symb:eocp}{type=symb,name={\ensuremath{e_{ocp}}},description={Fuel cell's open circuit voltage}}
\newglossaryentry{symb:eohm}{type=symb,name={\ensuremath{e_{ohm}}},description={Fuel cell's ohmic voltage loss}}
\newglossaryentry{symb:Ea}{type=symb,name={\ensuremath{E_a}},description={Activation energy}}
\newglossaryentry{symb:F}{type=symb,name={\ensuremath{F}},description={Faraday's constant}}
\newglossaryentry{symb:g}{type=symb,name={\ensuremath{g}},description={Specific Gibbs free energy}}
\newglossaryentry{symb:G}{type=symb,name={\ensuremath{G}},description={Gibbs free energy}}
\newglossaryentry{symb:h}{type=symb,name={\ensuremath{h}},description={Specific enthalpy}}
\newglossaryentry{symb:H}{type=symb,name={\ensuremath{H}},description={Enthalpy}}
\newglossaryentry{symb:k}{type=symb,name={\ensuremath{k}},description={Pre-exponential factor}}
\newglossaryentry{symb:KEq}{type=symb,name={\ensuremath{K_{Eq}}},description={Temperature dependent reaction equilibrium constant}}
\newglossaryentry{symb:Kp}{type=symb,name={\ensuremath{K_p}},description={Components’ activities dependent reaction equilibrium constant}}
\newglossaryentry{symb:Kt}{type=symb,name={\ensuremath{K_t}},description={Stodola's law flow coefficient}}
\newglossaryentry{symb:j}{type=symb,name={\ensuremath{j}},description={Current density}}
\newglossaryentry{symb:j0}{type=symb,name={\ensuremath{j_0}},description={Exchange current density}}
\newglossaryentry{symb:I}{type=symb,name={\ensuremath{I}},description={Current}}
\newglossaryentry{symb:M}{type=symb,name={\ensuremath{M}},description={Mass}}
\newglossaryentry{symb:n}{type=symb,name={\ensuremath{n}},description={Number of electrons exchanged in fuel cell oxidation reaction}}
\newglossaryentry{symb:S}{type=symb,name={\ensuremath{S}},description={Number of species in a fluid mixture}}
\newglossaryentry{symb:N}{type=symb,name={\ensuremath{N}},description={Number of volumes in a module}}
\newglossaryentry{symb:p}{type=symb,name={\ensuremath{p}},description={Pressure}}
\newglossaryentry{symb:Q}{type=symb,name={\ensuremath{Q}},description={Thermal power (positive entering)}}
\newglossaryentry{symb:R}{type=symb,name={\ensuremath{R}},description={Universal gas constant}}
\newglossaryentry{symb:Rohm}{type=symb,name={\ensuremath{R_{ohm}}},description={Ohmic resistance}}
\newglossaryentry{symb:T}{type=symb,name={\ensuremath{T}},description={Temperature}}
\newglossaryentry{symb:u}{type=symb,name={\ensuremath{u}},description={Specific internal energy}}
\newglossaryentry{symb:U}{type=symb,name={\ensuremath{U}},description={Internal energy}}
\newglossaryentry{symb:v}{type=symb,name={\ensuremath{v}},description={Specific volume}}
\newglossaryentry{symb:V}{type=symb,name={\ensuremath{V}},description={Volume}}
\newglossaryentry{symb:w}{type=symb,name={\ensuremath{w}},description={Mass flow rate}}
\newglossaryentry{symb:X}{type=symb,name={\ensuremath{X}},description={Mass fraction}}
\newglossaryentry{symb:Y}{type=symb,name={\ensuremath{Y}},description={Molar fraction}}

\newglossaryentry{greek:alpha}{type=greek,name={\ensuremath{\alpha}},description={Charge transfer coefficient}}
\newglossaryentry{greek:beta}{type=greek,name={\ensuremath{\beta}},description={Pressure ratio}}
\newglossaryentry{greek:gamma}{type=greek,name={\ensuremath{\gamma}},description={Heat transfer coefficient }}
\newglossaryentry{greek:lambda}{type=greek,name={\ensuremath{\lambda}},description={Homotopy operator parameter}}
\newglossaryentry{greek:rho}{type=greek,name={\ensuremath{\rho}},description={Density}}
\newglossaryentry{greek:tau}{type=greek,name={\ensuremath{\tau}},description={Thickness}}

\newglossaryentry{sub:des}{type=sub,name={\ensuremath{des}},description={On design value}}
\newglossaryentry{sub:i}{type=sub,name={\ensuremath{i}},description={i-th species of the considered mixture}}
\newglossaryentry{sub:in}{type=sub,name={\ensuremath{in}},description={Inlet quantity}}
\newglossaryentry{sub:j}{type=sub,name={\ensuremath{j}},description={j-th discretization volume of a model}}
\newglossaryentry{sub:k}{type=sub,name={\ensuremath{k}},description={Index of the considered homotopy level}}
\newglossaryentry{sub:n}{type=sub,name={\ensuremath{n}},description={Normalized with respect to nominal value}}
\newglossaryentry{sub:nom}{type=sub,name={\ensuremath{nom}},description={Nominal value}}
\newglossaryentry{sub:out}{type=sub,name={\ensuremath{out}},description={Outlet quantity}}
\newglossaryentry{sub:PEN}{type=sub,name={\ensuremath{PEN}},description={Positive-Electrolyte-Negative (referred to solid layers of the SOFC)}}
\newglossaryentry{sub:r}{type=sub,name={\ensuremath{r}},description={Chemical reaction index}}
\newglossaryentry{sub:ref}{type=sub,name={\ensuremath{ref}},description={Reference value}}
\newglossaryentry{sub:sat}{type=sub,name={\ensuremath{sat}},description={Saturation}}
\newglossaryentry{sub:tpb}{type=sub,name={\ensuremath{tpb}},description={Referred to fuel cell's triple phase boundary}}
\newglossaryentry{sub:v}{type=sub,name={\ensuremath{v}},description={Water vapor}}

\newglossaryentry{sup:el}{type=sup,name={\ensuremath{el}},description={Referred to fuel cell's electrodes}}
\newglossaryentry{sup:r}{type=sup,name={\ensuremath{r}},description={Reactant of reaction $r$}}

\makeglossaries

\begin{document}

\begin{frontmatter}



\title{Steady-State Initialization of Object-Oriented Advanced Thermal Power Generation System Models with Application to the Case of the SOS-CO\tsub{2} Cycle}


\author[deib]{Matteo Luigi De Pascali}
\ead{matteoluigi.depascali@polimi.it}
\author[deib]{Francesco Casella\corref{corr}}
\ead{francesco.casella@polimi.it}

\affiliation[deib]{organization={Politecnico di Milano, Dipartimento di Elettronica, Informazione e Bioingegneria},
	addressline={\\via Ponzio 34/5}, 
	city={Milan},
	postcode={20133}, 
	country={Italy}}

\cortext[corr]{Corresponding author}

\begin{abstract}
The forthcoming energy transition calls for a new generation of thermal power generation systems with low- or zero-emission and highly flexible operation.
Dynamic modelling and simulation is a key enabling factor in this field, as controlling such plants is a difficult task for which there is no previous experience and very short design times are expected. 
The steady-state initialization of those dynamic models is an essential step in the design process, but is unfortunately a difficult task which involves the numerical solution of large systems of nonlinear equations with iterative Newton methods, which is often prone to numerical failures.

In this work, several strategies and methodologies are discussed to successfully achieve steady-state initialization of first-principles equation-based, object-oriented models of advanced thermal power generation systems. 
These are presented in the context of the Modelica modelling language, but could be applied to other equation-based, object-oriented modelling and simulation environments.

Finally, the successful application of such strategies and methodologies to the SOS-CO$_2$ advanced power generation system is presented.
\end{abstract}

%
\begin{highlights}
\item Steady-state initialization of object-oriented advanced thermal power generation systems can be achieved reliably and efficiently resorting to the presented homotopy-based techniques.
\item Specific modelling techniques for components often employed to build advanced thermal power generation systems are presented.
\item A suite of open-source \emph{initialization blocks} to support the analysis, the simulation and the optimization of models of advanced thermal power generation systems is published on GitHub.
\item The challenging on- and off-design steady-state initialization of the SOS-CO$_2$ cycle model is successfully carried out, demonstrating the validity of the presented approach.
\end{highlights}


\end{frontmatter}



\glsaddall[types={symb,greek,sub,sup}]
\newpage
\printglossaries

\section{Introduction}
\label{sec:intro}
In the quest to achieve net-zero carbon emissions by the year 2050, which is the recommendation of the scientific community to avoid catastrophic outcomes caused by global warming \cite{IPCC2023} the power sector is currently undergoing a massive transition process, with dramatically increasing penetration of intermittent renewable power sources such as solar photovoltaic and wind power. 
Balancing electrical power system in these new scenarios will be increasingly challenging. 
In the final 2050 power systems configuration the balancing issues may be completely resolved by batteries, demand-side management, power-to-X and sector coupling, see, e.g., \cite{BogdanovEtAlAE2021}; however, during the transition process, there will be need for flexible on-demand power production to fill in the gaps left by renewable sources, which may be obtained by means of innovative, low- or zero-emissions thermal power generation units.

Among these innovative and clean energy systems, many still rely on combustion to convert the chemical energy of the fuels and may feature revamped traditional components (e.g. compact and very efficient heat exchangers and innovative turbomachineries), unconventional components (e.g. fuel cells), and unconventional working fluids, e.g., organic mixtures, hydrogen, or carbon dioxide.
The operating conditions of these cycles are often also extreme, with unprecedented temperatures and pressures employed to take advantage of the working fluid's thermodynamic properties.
A key feature for new thermal power generation systems operating in a high renewables penetration scenario is flexibility: being able to ramp up and down the power output within a few minutes, as well as being able to keep the plant running at near-idle conditions, ready to kick in quickly as required without performing shut-down and start-up procedures, both help achieving grid stability and increasing the revenues on the power markets.

In this context, building dynamic models of such plants as early as possible in the design process allows to assess how flexible a proposed new design is and provides even more valuable information for plant concepts which are still on the drawing board and for which there is no prior operational experience. 
Preliminary designs might be modified to improve the plant's behaviour in off-design and transient conditions, based on the outcome of such dynamic simulations.

Steady-state initialization is a crucial issue when dealing with dynamic models of thermal power generation systems and is required in order to compute typical simulation scenarios, in which the power generation system starts from a stable operating condition and then reacts to some variation in its inputs, such as changing the power output set point or reacting to grid frequency variations because of primary frequency control.
Furthermore, when tackling the study of innovative plant designs, it is often the case that only on-design operating conditions at full load are initially known; once a dynamic model of the plant has been built, it is very convenient to be able to re-use it to study off-design operating strategies, e.g. for part-load operation or operation with off-design boundary conditions such as high cooling temperature for condensers; this again requires computing steady-state solutions of the dynamic model.

One approach which is often used to obtain such steady-state conditions is to run relaxation transients, in which the design values of the system inputs are fixed, the plant model state variables are initialized with some guess values and then a transient simulation is run until a steady-state is achieved. This approach suffers from three major shortcomings. 
The first is that such a transient simulation could be very time consuming and even fail if the initial states have not been initialized properly.
The second is that if the sought-after equilibrium point of the plant is not asymptotically stable, then the simulation will actually never approach it. 
The third, which is particularly relevant when initializing in off-design conditions, is that the values of the system inputs to obtain the required steady-state outputs are unknown. 
Addressing the second and third shortcomings requires to add some kind of feedback control system to the model, in order to stabilize the system and steer its outputs to the required values. 
This can be problematic, because during the initial analysis of off-design operating strategies for a new type of plant, a control system design is not yet available. 
Furthermore, the design of control systems often requires to run preliminary tasks, such as computing the linearized dynamics of the plant around certain operating points, that require the steady state operating point to be known, ending up in a catch-22 situation.

The alternative approach is to directly solve the dynamic system equations for steady-state conditions, by adding initial equations that set all state derivatives to zero and prescribe the desired input or output values. 
As it is well-known among practitioners, these systems of equations can be very large and strongly nonlinear, so that solving them reliably with iterative methods such as Newton-Raphson's can be a very challenging proposition.

Although the steady-state initialization problem for thermal power generation systems must somehow be tackled whenever a dynamic model is used, there are surprisingly few (if any) papers in the literature that try to address this problem in a systematic way.  
Papers reporting the outcome of studies involving dynamic models often show results starting from steady-state, but never get into the details of how they were computed, as if this was some kind of menial task to be typically carried out by PhD students as a part of their training, and not worth discussing in serious publications. 
The second author of this paper has been dealing with this problem for quite some time, specifically in the context of dynamic models written with the Modelica language \cite{mattssonPhysicalSystemModeling1998}; in this area, he published some early recommendations on the use of homotopy in \cite{CasellaSielemannSavoldelli2011}.

The main aim of this paper, which builds upon those earlier result and on more recent research work, is thus to present a \emph{systematic} approach to address the steady-state initialization of equation-based dynamic models of thermal power generation system in a reliable and sound way. The proposed methodology is implemented in the Modelica language, although the underlying concepts could in principle be used in any a-causal, equation-based modelling language. 

These concepts are then successfully demonstrated in a practical, and quite challenging, use case, namely the steady-state initialization of the Solid-Oxide Semi-Closed CO$_2$ (SOS-CO$_2$) cycle \cite{scaccabarozziSolidOxideSemiclosed2021}, to illustrate the validity of the proposed methodology. 
The SOS-CO$_2$ is an innovative thermal power generation system that combines a recuperated Brayton cycle, working between 0.1 and 3 MPa, with an internal reforming solid oxide fuel cell (SOFC), placed before the combustor, where a CO$_2$-O$_2$ mixture replaces air. 
This cycle represents a really tough challenge for steady-state initialization, since it combines the highly nonlinear behaviour of the real gas mixtures, of the chemical and electrochemical reactions that take place in the fuel cell, and of the complex thermal interactions taking place in 1D heat exchanger models. 
Furthermore, the previously available results from \cite{scaccabarozziSolidOxideSemiclosed2021} only provided on-design values for the main operating variables, leading to the requirement to use the dynamic model both for the study of off-design operating strategies and for the design of the control system.

The main outcomes of this work are systematic guidelines on how to write Modelica models of thermal power generation systems that can be successfully initialized in steady-state, both in on- and off-design conditions. 
Some of these guidelines are actually embedded in a small, open-source Modelica library, that can used out-of-the-box to support steady-state initialization in a variety of configurations: on- and off-design, open-loop and closed-loop, and, last but not least, when exporting the model as an FMU according to the FMI standard \cite{FMI}.

The paper is organized as follows: Section~\ref{sec:whyModelica} presents the employed Modelica modeling framework; Section~\ref{sec:strategies} introduces the strategies that can be used to ease the convergence of the steady-state initialization problems for thermal power generation system models; these strategies are then applied to individual component models and to whole plant assembled from the individual components in  Section~\ref{sec:ModelicaPlantInit}; Section~\ref{sec:sosco2} describes the application of these concepts to the use-case of the SOS-CO$_2$ cycle steady-state initialization; Section~\ref{sec:conclusions} draws the conclusion for this work.

\section{The modelling framework}
\label{sec:whyModelica}
The authors of \cite{CasellaLevaMCMDS2006} list a number of desirable features for a modeling and simulation environment suitable for thermal power generation systems. 
A key identified feature of a modeling software is providing support for initialization of dynamic simulations.
If this crucial step is not successfully completed, the model itself serves no purposes, since the simulation cannot be started.
For thermal power generation systems this task is particularly difficult.
\begin{itemize}
	\item The number of equations required to properly describe such systems may reach or even exceed one hundred thousand, possibly requiring to solve very large core systems of nonlinear equations during initialization, with possibly thousands of unknowns.
	\item Initial values for the states variables in thermal power generation systems are likely to be unknown: conversely to what happens for mechanical systems in which initial positions, velocities and accelerations are often known a priori, exact values for temperatures, pressures, specific enthalpies, and compositions distributions in energy systems (which are typically chosen as state variables) are often unavailable before completing a first simulation.
	\item Proper initial guess values of iteration variables for the nonlinear initialization equations might be unavailable or not sufficiently accurate, possibly leading to Newton-Raphson algorithm convergence failures (as for the point above, also enthalpies and densities distributions usually are unknown in the components).
\end{itemize}

Innovative thermal power generation system models often check all these points, making their initialization difficult to achieve.
In particular, the second point of the list forces users to resort to steady-state initialization instead of fixed state variables initialization. The steady-state initialization problem can be subject to structural analysis to solve the equations as much as possible in a sequential way. 
Unfortunately, the structure of such problems is usually characterized by the presence of one, very large implicit system of equations (corresponding to a strong component in the equation dependency graph), which couples the behaviour of all the system components. Providing good initial guesses for the solver of such system is non-trivial, tedious, and often frustrating task, if the iterative solver then fails.

Hence, a methodology that offers useful insights about the initialization steps, that points to the causes of convergence failures and offers techniques to ease the initialization task (like the ones presented in \cite{CasellaSielemannSavoldelli2011} and \cite{casella2021choice}) should be adopted to approach such tasks.

The Modelica language \cite{mattssonPhysicalSystemModeling1998} is a powerful modelling and simulation tool that satisfies the requirements listed in \cite{CasellaLevaMCMDS2006}; following the freely available Modelica language specification, many simulation environments can compile executable code for simulation, under both open source or commercial licenses.
Since Modelica was first made available to the public, the size and complexity of the models created in this framework that can be handled by state-of-the art tools has grown considerably; many power plants model were coded and published, going from combined cycle power to nuclear reactors, parabolic trough collectors and supercritical cycles  \cite{casellaFastStartupCombinedCycle2006,CammiEtAlModelica2005,eckDynamicsControlParabolic2007,CasellaColonnaSCO2Cycle2011}.
Specific libraries for this kind of systems have been also published with open-source licenses \cite{ClaRa,ThermoSysPro, ThermoPower}.

For these reasons, Modelica was selected as the underlying framework to implement the methodologies proposed in this paper. Nonetheless, these strategies are based on general mathematical and physical principles and rely on Newton-Raphson solvers for the resolution of systems of equations, so they should be applicable to any equation-based, object-oriented modelling environment.

\section{Strategies for direct steady-state initialization}
\label{sec:strategies}
As discussed in \cite{CasellaBachmannAMC2021}, strongly nonlinear equations in the initialization problem are more likely to be the cause of convergence failure, if the initial guesses of the variables appearing nonlinearly in them are not close enough to the solution. Additionally, it is well-known to practitioners that the larger a nonlinear equation system is, the more likely it is to have convergence problems if the initial guess is not close enough to the solution, although, up to the authors' knowledge, there is no published theoretical result backing this statement.

A first strategy that is useful to avoid convergence failures when solving nonlinear initialization problems is to make sure that the nonlinear equations passed to the Newton-Raphson solver are properly scaled \cite{CasellaBraunEOOLT2017}, i.e., both unknowns and equation residuals should have the order of magnitude of unity. 
This is particularly important when SI units are employed in the models, as it is standard practice in Modelica, because the order of magnitude of the variables and of the residuals may span an interval that can be larger than the machine precision. 
Modelica tools can automatically perform such scaling, using the \emph{nominal} attribute of each variable as a scaling factor for the unknowns and automatically deriving the scaling factor of residuals via sensitivity analysis, as explained in \cite{CasellaBraunEOOLT2017}. 
It is then of the utmost importance that the variables of the developed Modelica models all have properly set nominal attributes.

A second strategy that can be quite effective to avoid convergence problems is the use of problem-specific homotopy transformations on key nonlinear equations \cite{CasellaSielemannSavoldelli2011}. 
Homotopy is a linear transformation between two expressions, which is used by Modelica environments during initialization \cite{SielemannEtAlModelica2011}.
\begin{equation}
	y(x) = \lambda \cdot f(x) + (1-\lambda) \cdot g(x);
	\label{eq:homotopy}
\end{equation}
the idea is that some expression $f(x)$ in the actual model is replaced by $y(x)$, which is a convex linear combination between $f(x)$ and a suitably simplified version $g(x)$ of it.

The initialization problem is first solved with $\lambda = 0$; the idea is to pick the simplified expression $g(x)$ to ease the convergence of the solver. 
Once the simplified model is solved, the solver gradually increases $\lambda$, thus solving a sequence of nonlinear problems that progressively approximate the actual one better and better, until reaching $\lambda = 1$. 
By taking small enough steps for $\lambda$, each problem in this sequence can be made close enough to the previous one so as to guarantee convergence, if the solution of the previous step is used as an initial guess for the current one, eventually leading to the solution of the original problem in a kind of bootstrapping process.

It is important to note that, although the homotopy expression \eqref{eq:homotopy} changes continuously, the corresponding solution of the nonlinear system may incur into singularities \cite{SielemannEtAlModelica2011}, which take place whenever $\dv{x}{\lambda} \to \infty$, where $x$ is any of the problem's variables. 
As a heuristic rule, this can be avoided if the behaviour of the simplified system is not too qualitatively different from the behaviour of the actual one. 
If there are no singularities during the homotopy transformation process, the success of the solution of the initialization process ultimately only depends on whether the \emph{simplified} initialization problem can be solved reliably or not. 

It is then crucial to understand which principles and rules should be followed to introduce simplifications that maximize the likelihood of convergence for the simplified problem at $\lambda = 0$, while reducing or even completely eliminating the need of setting accurate initial guesses for its unknowns. 
This can be achieved by following two principles.

The first principle is to introduce simplifications that replace strongly nonlinear equations with less non-linear or even fully linear ones at $\lambda = 0$, since this decreases the dependency of the convergence success on the accuracy of the initial guesses for the Newton-Raphson solver, or even eliminates it completely in the case of linear equations, as explained in \cite{CasellaBachmannAMC2021}.

The second principle is to introduce simplifications at $\lambda = 0$ that eliminate critical couplings between the equations of the original initialization problem, so that the solution of the simplified model can be automatically split by the Modelica tool into the sequential solution of several smaller problems. 
In structural analysis terms, the idea is to remove the dependency of some equations on certain variables, so that a large strong component in the initialization equation dependency graph can be split into several smaller strong components. 
This has several beneficial outcomes:
\begin{itemize}
  \item the computational load is reduced;
  \item the split systems are smaller and less nonlinear, so they are much more robust with respect to convergence of Newton-Raphson's method;
  \item there is no need to worry about the initial guesses of a lot of unknowns of the original large, coupled system of equations, because their values are automatically computed by the solution of the previous strong components in the solution sequence. 
\end{itemize}

For example, it is often the case that mass balance, energy balance and chemical reaction equations are coupled together by the component's physical behaviour, leading to a large, coupled nonlinear initialization problem. 
If some simplifications can be introduced that break this coupling, then the mass balance equations can be first solved separately and more reliably, allowing to compute the mass flow rates and pressures, whose values will then be used to solve the energy balance and chemistry equations more reliably.

The second principle described above of course only works if the Modelica tool employed to generate the simulation code applies structural analysis and simplifications specifically to the initialization problem at $\lambda = 0$, as suggested in \cite{CasellaSielemannSavoldelli2011}. 
Up to the author's knowledge, the commercial tool Dymola and the open-source tool OpenModelica actually do that.

As a final remark, the proposed approach has two very nice features, as will become apparent from the next three Sections. 
The first is that the simplifications required to successfully solve the initialization problem only need to be applied to a few selected equations. 
The second is that, although the choice of effective simplifications requires full awareness of the mathematical structure of the underlying initialization problem (which is not trivial), once the right simplifications have been selected the actual process of splitting the original initialization problem into a sequence of smaller problems is carried out automatically by the Modelica tool, without any need to program it explicitly.

\section{Individual component models}
\label{sec:components}

In this section, models for components often employed to build advanced thermal power generation systems are presented, applying the strategies discussed in the previous Section to ease the solution of steady-state initialization of system models using them.
Particular attention will be devoted to the equations that are most critical for the steady-state plant initialization problem. 
The presented component models are all relevant for the SOS-CO$_2$ cycle, but could also find use in a wide range of advanced thermal power generation systems; they can be considered as paradigmatic example cases of strategies to ease the steady-state initialization of thermal power generation systems, and can be used as a source of inspiration for other types of components that are not explicitly mentioned here.

\subsection{Turbine models}
Turbines employed in advanced thermal power generation systems often rely on high expansion ratios to maximize the specific work extracted by such expanders.
The inlet temperature of such components might vary greatly and cooled turbines are as common as uncooled ones.
Thus, in case of higher inlet temperatures, multiple cooling flows could be sent in the first turbine stages and heating flows might be extracted as well for preheating purposes. 
Splitting the expansion among multiple fictitious stages and injecting/extracting flows through blocks that impose a custom link between flows and pressures might be a straightforward solution for these cases.

Considering the size of the component and the pressure ratio at which it operates, the residence time of the flowing fluid can be neglected and the expander might be modeled as 0-D object.
The behaviour of this component is described through the Stodola's ellipse law \cite{cookePredictionOffDesignMultistage2015}, as described in Equation \eqref{eq:stodola}
\begin{equation}
	w = K_t \cdot \sqrt{p_{in}\cdot \rho_{in}} \cdot \sqrt{1-\left (\frac{1}{\beta}\right )^2}
	\label{eq:stodola}
\end{equation}
Tuning the parameter $K_t$ and the turbine's isentropic efficiency allows to match the on-design desired performance and regulates the off-design behaviour according to changes in boundary temperatures and pressures.
Given the presence of strong nonlinearities introduced by the square root and the multiplications, it is convenient to simplify this expression as in Equation~\eqref{eq:simplifiedStodola} and use it as the simplified term with the homotopy operator.
\begin{equation}
	w = \frac{w_{nom}}{p_{nom}}\cdot p_{in}
	\label{eq:simplifiedStodola} 
\end{equation}
where nominal values refer to on-design, full load operating conditions. 
Equation~\eqref{eq:simplifiedStodola} is obtained as shown in the equations below, considering that $\frac{1}{\beta^2} \rightarrow 0$ in the cycle operating conditions and that the inlet temperature does not change considerably in partial load conditions, and that the gas behaviour can be approximated by the ideal gas law:
\begin{equation}
	w = K_t \cdot \sqrt{p_{in}\cdot \rho_{in}} \cdot \sqrt{1-\left (\frac{1}{\beta}\right )^2} \approx 
	K_t \cdot \sqrt{\frac{1}{RT_{in}}} \cdot p_{in} \approx C \cdot p_{in},
\end{equation}
where the constant $C$ is then determined based on the available on-design operating conditions.
This very simple linear equation removes at $\lambda = 0$ the link between the fluid dynamic phenomena and the thermal ones, neglecting the dependency on the temperature through the density on the mass flow rate.

Besides getting rid of the nonliner behaviour of Equation \eqref{eq:stodola}, Equation \eqref{eq:simplifiedStodola} allows to decouple the mass flow rate calculation at $\lambda = 0$ from the power output calculation, which is influenced both by the mass and energy balances of the components upstream and involves many more equations.

\subsection{Heat Exchanger models}
\label{sec:hex}
Heat exchangers are key components of almost all thermal power generation systems, in particular advanced ones that try to improve the overall efficiency; examples include recuperators in regenerative gas cycles, economizers, evaporators, and superheaters in Rankine cycles, condensers, etc. 

The heat exchange units can be modeled by the series connection of modules, each containing  several finite discretization volumes (index $j$ in the following equations) describing the fluid flow through the hot and the cold sides, as well as the heat storage in the in-between walls. 
This division in modules and volumes allows to refine the temperature, pressure and composition grids as needed: finite volumes are characterized each by a temperature and composition state, while the pressure state is unique for each module and shared among its volumes, with pressure losses lumped at the boundaries.

Since the pressure losses through heat exchangers used in power generation units are typically low (a few percentage points of the absolute pressure), decreasing the number of pressure states is a reasonable assumption that helps reducing the model complexity; also, having one pressure state for each volume would lead to very stiff differential equations, due to the very small pressure loss between two adjacent volumes, so lumping the pressure of several volumes together is beneficial. 
On the other hand, the price to pay for this reduction is that the equations to compute state derivatives of each entire module during simulation become coupled and need to be solved by a nonlinear solver, which suggests limiting the number of volumes in a module to avoid the need of solving large nonlinear equations at each time step of the simulation. 
This means that a suitable balance between these two aspects should be found.

The equations that describe the heat transfer process between the channels and the wall of a single module are the following.
\begin{align}
	& \dv{M_j}{t} = w_{j,in} - w_{j,out}, & j = 1 \ldots N
	\label{eq:massBalHX} \\
	& \dv{M_jX_{i,j}}{t} = w_{j,in}X_{i,j,in}-w_{j,out}X_{i,j,out}, & i = 1 \ldots S, \quad j = 1 \ldots N
	\label{eq:compBalanceHX}\\
	& \dv{U_j}{t} = w_{j,in}h_{j,in} - w_{j,out}h_{j,out} + Q_j, & j = 1 \ldots N
	\label{eq:enBalanceHX} \\
	& Q_j = \gamma_j \cdot S_j \cdot (T_{j,wall} - T_{j,flow}), & j = 1 \ldots N
	\label{eq:convThermalPower}\\
	& \gamma_j = \gamma_{j,nom}\cdot \left(\frac{w_j}{w_{j,nom}}\right)^{0.8} \cdot \left(\frac{p_j}{p_{j,nom}}\right )^{0.5}, & j = 1 \ldots N
	\label{eq:heatTransferCoeff}
\end{align}

Equation~\eqref{eq:massBalHX} is the total mass balance, while \eqref{eq:compBalanceHX} is the mass balance referred to each species of the fluid mixture.
The energy balance in Equation~\eqref{eq:enBalanceHX} is coupled with Equation~\eqref{eq:convThermalPower}, which defines the thermal power exchanged between the fluid in the channels and the wall separating them.
Since the purpose of such models is to carry out system-level plant-wide studies, Equation~\eqref{eq:heatTransferCoeff} can be adopted to define the convective heat transfer coefficient when only on-design values are known.
This is a simplification taking place of more complex formulations that require the evaluation of Reynolds and Prandtl numbers, as in the case of the Dittus-Boelter or Gnielinski correlations.

Equations \eqref{eq:massBalHX}, \eqref{eq:compBalanceHX} and \eqref{eq:enBalanceHX} can be tricky to solve during transient simulations if they are coded as presented above and several strategies can be adopted to prevent convergence issues.
First and foremost, pressure, temperatures and composition should be used as state variables, since most equations used during simulation, when such state values are known at each step, can be solved explicitly if these variables are known. 
Hence, equations \eqref{eq:dMdt} and \eqref{eq:dUdt} are introduced to change the state variables, as recommended, making $\dv{M_j}{t}$ and $\dv{U_j}{t}$ dummy derivatives. 
Moreover, to remove the total mass from the derivative operator, Equation~\eqref{eq:compBalanceHX} can be reformulated as in Equation~\eqref{eq:compBalanceHX_V2}, by taking into account the total mass balance Eq~\eqref{eq:massBalHX}.

\begin{align}
	& \dv{M_j}{t} = -V_j \cdot \rho_j ^ 2 \cdot \left(\pdv{v_j}{T_j} \cdot \dv{T_j}{t} + \pdv{v_j}{p_j} \cdot \dv{p_j}{t} + \sum_{i=1}^{S}{\pdv{v_j}{X_{i,j}} \cdot \dv{X_{i,j}}{t}}\right) 
	\label{eq:dMdt}\\
	& \dv{U_j}{t} = M_j \cdot \left(\pdv{u_j}{T_j} \cdot \dv{T_j}{t} + \pdv{u_j}{p_j} \cdot \dv{p_j}{t} + \sum_{i=1}^{S}{\pdv{u_j}{X_{i,j}} \cdot \dv{X_{i,j}}{t}}\right) + \dv{M_j}{t} \cdot u_j 
	\label{eq:dUdt} \\
	& M_j\cdot\dv{X_{i,j}}{t} = w_{j,in}\cdot\left(X_{i,j,in} - X_{i,j,out}\right) 
	\label{eq:compBalanceHX_V2}
\end{align}

While the formulation above is beneficial during transient simulation, it introduces other possible critical issues at initialization, which should be addressed.
First, the new expression proposed in Equation~\eqref{eq:compBalanceHX_V2} can possibly cause problems when considering steady-state initialization: when the initial equations $\frac{dX_{i,j}}{dt} = 0$ are added, the Modelica tool can infer that the left-hand-side of \eqref{eq:compBalanceHX_V2} is zero, thus replacing it with

\begin{equation}
	0 = w_{j,in}\cdot\left(X_{i,j,in} - X_{i,j,out}\right)
	\label{eq:initCompo2}
\end{equation}

Now, this equation has two solutions: one is $X_{i,j,in} = X_{i,j,out}$, the other is $w_{j,in} = 0$. 
Of course the first is the relevant one for the problem at hand, and it would allow to considerably simplify the solution of the system, because it allows to remove all the compositions from the set of unknowns of the nonlinear system. 
However, the tool doesn't know this, so it will keep all the nonlinear equations \eqref{eq:initCompo2} and their unknown compositions in the nonlinear system, making it unnecessarily much larger and much more difficult to solve. 
It is therefore recommended to write the initial equation explicitly as
\begin{equation}
	X_{i,j,out} = X_{i,j,in}
	\label{eq:initCompo}
\end{equation}
so that the tool will be able to remove the compositions from the unknowns of the implicit nonlinear systems of equations.
Unfortunately, now that Equation~\eqref{eq:initCompo} is employed as steady-state initial equation for the compositions, the state derivatives $\dv{X_{i,j}}{t}$ are not explicitly assigned to zero, but (unnecessarily!) remain as unknown variables of the implicit nonlinear initialization system. 
Furthermore, the lack of an explicit $\dv{X_{i,j}}{t} = 0$ equation prevents to perform a chain of symbolic simplifications that can substantially simplify the nonlinear equations for initialization. 

Explicitly stating that $\dv{X_{i,j}}{t} = 0$, together with the other initial equations $\dv{T_j}{t}=0$, $\dv{p_j}{t} = 0$, would allow the Modelica tool to symbolically infer from \eqref{eq:dMdt} that $\dv{M_j}{t} = 0$ , and hence to infer that $w_{j,in} = w_{j,out}$ from Equation~\eqref{eq:massBalHX}. 
These symbolic manipulations would allow to remove $\dv{X_{i,j}}{t}$, $\dv{M_j}{t}$ and $w_{j,out}$  from the unknowns of the nonlinear system, as well as removing Equations \eqref{eq:dMdt} and \eqref{eq:dUdt}, whose residuals are extremely cumbersome to compute, from the set of the nonlinear equations. 
This would lead to a much smaller and much less nonlinear system, whose convergence is much faster and robust. 
The lack of such an explicit $\dv{X_{i,j}}{t} = 0$ equation in the initialization problem thus prevents a strategic and massive symbolic simplification of the nonlinear equations to be solved, leading to a formulation that is bound to fail to converge for even moderately sized systems.

This problem can be easily solved in Modelica by tweaking the partial mass balance Equation \eqref{eq:compBalanceHX_V2} so that during steady-state initialization it explicitly states that $\dv{X_{i,j}}{t} = 0$:

\begin{align}
	\begin{split}
		&\text{if initial() then}\\
		&\quad \dv{X_{i,j}}{t} = 0;\\
		&\text{else}\\
		&\quad M_j\cdot\dv{X_{i,j}}{t} = w_{j,in}\cdot\left(X_{i,j,in} - X_{i,j,out}\right)\\
		&\text{end if;}
		\label{eq:initStratComp}
	\end{split}
\end{align}
Note that the structural simplification brought by Equation \eqref{eq:initStratComp} does not explicitly involve homotopy, so it is valid throughout the entire homotopy transformation, leading to smaller nonlinear systems also for $\lambda \neq 0$, substantially reducing the computation time of the homotopy transformation.

\subsection{Intercooler models}
Intercoolers are usually placed at the discharge of compressors to cool down the compressed flow and reduce the specific work of the trailing turbomachineries.
There are many possibilities to model these components:
\begin{itemize}
	\item directly link the temperatures of the hot and cold fluid through $\varepsilon\!-\!\text{NTU}$ formulas, completely neglecting the effects of mass and energy storage;
	\item use full heat exchanger models as those shown in the previous sub-section, with cold water or air as the cold fluid;	
	\item employ more simplified equations.
\end{itemize}

The first option introduces highly nonlinear equations involving the logarithmic mean temperature difference, which could complicate the solution of the associated nonlinear equations during initialization. 
The second option involves the same considerations made for the heat exchangers; the problem in this case is that using full-detail heat exchanger models for such components could substantially increase the size and computational load of the whole plant model, which may not be worth it, given the modest impact that the intercoolers play in the overall plant behaviour. 

Since the intercoolers thermal load is usually much smaller than the one of the main heat exchangers that provide heat to the working fluid from some thermal source to produce power, a simplified model can be employed to avoid the downsides of the aforementioned strategies. 
Hence, in many cases, the outlet temperature of the working fluid can be ideally set to a fixed value, as if the flow of water is regulated to obtain the same temperature at all loads. 
In case the inner volume of the working-fluid side is significant, one can model such heat exchangers as 0D volumes, with a dynamic mass balance equation and with the energy balance equations replaced by an equation fixing the outlet temperature; this allows to take into account the variable mass storage inside the volume as the pressure changes. 
Since the model purpose is to conduct plant-wide studies to ultimately tune a controller for the plant and test its flexibility, this degree of approximation can be considered acceptable in most cases.

In both cases, fixed outlet temperature and 0D model with fixed outlet temperature, solving the initial equations is trivial, since the temperature is fixed, the outlet flow rate is equal to the inlet flow rate, and the outlet composition is equal to the inlet composition.

\subsection{Vapour fraction condenser models}
Oxy-combustion cycles such as SOS-CO$_2$ use pure oxygen as an oxidant for the combustion process, instead of air. 
Ideally, the product of combustion is a mixture of CO$_2$ and water vapour, which needs to be removed by condensation before sending the CO$_2$ flow to permanent underground storage. 
Such condensers can be modeled as 0-D flash tanks connected at the discharge of each intercooler, that ideally separate the two fluid phases through saturation pressure evaluation.

The fraction of liquid water is evaluated with Raoult's law and the saturation pressure is computed by a fitting polynomial as in equations \eqref{eq:raoult} and \eqref{eq:psat}.
\begin{align}
	& p_{sat} = Y_v \cdot p_{in}
	\label{eq:raoult} \\
	& p_{sat} = \exp(a_4 + T \cdot (a_3 + T \cdot (a_2 + T \cdot a_1)))
	\label{eq:psat}
\end{align}

Equation \eqref{eq:psat} evaluates the polynomial using Horner's rule \cite{burrus2003horner}. 
This is a practice which should be adopted systematically for polynomial expressions evaluation in equation-based models, since it guarantees to perform the minimum number of operations and, most importantly, guarantees the maximum numerical robustness when high-order polynomials are used.

The steady-state initialization strategy for such condensers is to introduce homotopy-based simplifications that aim at splitting the computation of fluid dynamics, thermal, and chemical phenomena at $\lambda = 0$, as discussed in Section \ref{sec:strategies}. 
In the full model, these phenomena are coupled together by the influence of thermal and chemical effects on the condensation rate, thus on the mass balance, so their reciprocal dependencies result into large and strongly nonlinear strong components in the initialization problem.

In the condenser models, this coupling can be broken in a very simple way by neglecting the outgoing condensed water flow rate in the total mass balance equation when $\lambda = 0$; this leads to a trivial steady-state mass balance equation $w_{out} = w_{in}$. 
In this way, the original dependence of the out-going flow rate on the (thermal-chemical) condensation phenomena is removed and the mass balance equations get decoupled from the energy- and chemistry-related phenomena, allowing to solve them first separately for the flow rates and pressures. 
Since the total condensed water stream is much lower than the overall working fluid flow rate (a few percentage points at most), the results obtained with this simplification at $\lambda = 0$ are very close to the actual ones, thus representing an ideal choice of simplified model for the homotopy process.

\subsection{Fuel cell models}
Advanced systems may resort on alternative, more efficient ways to oxidize the fuel supplied, without employing combustion.
Hence, fuel cells are often included to build hybrid-systems to convert the chemical energy stored in the fuels into electrical energy.
In this paper we discuss steady-state initialization strategies for a specific Solid Oxide Fuel Cell (SOFC) model, presented in detail in \cite{depascaliControlOrientedModelica1D2023}. 
Similar concepts could be applied to other fuel cell models, also of different nature, e.g., ambient-temperature Proton Exchange Membrane (PEM) fuel cells.

The core of the considered fuel cell model is a 1-D model representing a single cell, with a pair of anodic and the cathodic channels, a membrane that separates the two flows, and interconnecting plates confining the fuel and the oxidant in the channels and carrying the generated electrical current.
Thanks to the assumption that the adjacent channels of a fuel cell work in similar conditions, entire SOFC modules built by putting many of these cells in parallel are modelled by multiplying the extensive variables of an assembled channel couple by the total number of channels.

This component is the hardest to initialize among the ones presented in this Section, since it exploits highly nonlinear, coupled thermal, chemical and electro-chemical phenomena to convert the fuel chemical energy into electrical power. 
Moreover, the 1D process must be discretized in suitable number of discrete volumes (e.g., 10) along the axial dimension, resulting in large systems of coupled equations. 
For each discretization volume in the 1D cell model, Equations \eqref{eq:voltageFuelCell}-\eqref{eq:tpbPressure} can be written, which contain logarithms, exponentials and square roots; all variables in those equations are specific to each $j$-th volume, except for the cell voltage $E$, which is unique for the entire cell thanks to the conducting plates and thus couples the equations of all volumes together:

\begin{align}
	& E = E_{ocp} - e_{ohm} - e_{conc} - e_{act}^{cathode} - e_{act}^{anode} 
	\label{eq:voltageFuelCell}\\[10pt]
	& E_{ocp} = -\frac{\Delta g_{0,hor}}{nF} - \frac{RT}{nF}\cdot \ln{ \left(\frac{p_{H_2O}}{p_{H2}\cdot \sqrt{\frac{p_{O2}}{p_{ref}}}}\right)} 
	\label{eq:vOCP}\\[10pt]
	& e_{ohm} = R_{ohm} \cdot j 
	\label{eq:ohmicLoss}\\[10pt]
	& e_{conc} = \frac{RT}{2F} \cdot \qty(\ln\qty(\frac{p_{H_2O,tpb}}{p_{H_2O}} \cdot \frac{p_{H_2}}{p_{H_2,tpb}}) + 0.5\cdot\ln\qty(\frac{p_{O_2}}{p_{O_2,tpb}}))  \\[10pt]
	& e_{act}^{el} = \frac{RT}{\alpha nF}\ln\left(\frac{j}{2j_0^{el}}+\sqrt{\left(\frac{j}{2j_0^{el}}\right)^2+1}\right),  \quad\text{\footnotesize{el}} \in \text{\footnotesize{\{anode, cathode\}}} 
	\label{eq:v_act_explicit}\\[10pt]
	& j_0^{el}=\frac{RT_{PEN}}{nF}k^{el}\cdot\exp\left(-\frac{E_{a}^{el}}{RT_{PEN}}\right), \quad \text{\footnotesize{el}} \in \text{\footnotesize{\{anode, cathode\}}} 
	\label{eq:j0}\\[10pt]
	\begin{split} & p_{i,tpb} = p - (p-p_{i}) \cdot \exp\qty(RT \cdot \frac{\tau^{el}}{4FD_{Eff,i}, \,\,p^{el}}\cdot j), \\[4pt]
	& \quad\quad\quad \text{\footnotesize{i}} \in \text{\footnotesize{\{H$_2$O, H$_2$, O$_2$\}}}, \text{\footnotesize{el}} \in \text{\footnotesize{\{anode, cathode\}}} 
	\label{eq:tpbPressure}
	\end{split}
\end{align}

Equation \eqref{eq:voltageFuelCell} links the overall cell voltage $E$ to the open circuit voltage $E_{ocp}$ and to three kinds of voltage losses, which depend on the extracted current $j$, which is different for every discretization volume even if the overall voltage, the left-hand side term, is equal for all the volumes.
Hence, to evaluate the steady-state distribution of the current along the axial direction, a huge coupled nonlinear system of strongly nonlinear equations must be solved if no simplifications are adopted. 
Experience quickly showed that getting successful convergence of the Newton-Raphson algorithm on such a problem is an elusive goal, particularly if a non-trivial number of discretization volumes is chosen.

To resolve this issue, the homotopy method was used, introducing several strategic simplifications at $\lambda = 0$.

The first one is to approximate Equation \eqref{eq:voltageFuelCell} with a linear expression of the form $E=aI+b$, with suitable constant coefficients $a$ and $b$ that somehow approximate the fuel cell polarization curve. 
In this way, the electrical part of the system becomes linear and decoupled from the chemical part, so it can be solved easily even in case of large $N$, without the need of initial guess values, obtaining the values of the currents for each volume as inputs for the electro-chemical equations. 
Later on, through the homotopy transformation, the approximated linear polarization curve is gradually morphed into the real one, which depends on the thermo-chemical phenomena described by Equations (\ref{eq:voltageFuelCell})-(\ref{eq:tpbPressure}).

Regarding the voltage losses, their expressions can be simplified at $\lambda = 0$ to also decouple thermal and electro-chemical phenomena.
First, the direct influence of the fluid temperature $T$ in Equations \eqref{eq:vOCP} -\eqref{eq:tpbPressure} can be removed by approximating it with its constant nominal value $T_{nom}$ at $\lambda = 0$.
Equation \eqref{eq:vOCP} can be simplified at $\lambda = 0$ by replacing the logarithmic term with a reasonable fixed value; furthermore, the $\Delta g_{0,hor}$ term may be approximated by considering a fixed nominal temperature $T_{nom}$ while integrating over $T$ Van 't Hoff's law shown in Equation~\eqref{eq:vanthoff}.

\begin{equation}
	\dv{\ln K_{Eq,r}}{T} = -\dv{\frac{\Delta G_r^0(T)}{RT}}{T} = \frac{\Delta H_r^0(T)}{RT^2}
	\label{eq:vanthoff}
\end{equation}

Regarding the activation losses, the original equation to determine $e_{act}^{cathode}$, $e_{act}^{anode}$ is the Butler-Volmer equation
\begin{equation}
	j \!= \!j_0^{el}\!\!\left[\exp\!\left(\frac{\alpha nF}{RT_{PEN}}e_{act}^{el}\right)\!\!-\!\exp\!\left(\!\!-\frac{\left(1\!-\!\alpha\right)nF}{RT_{PEN}}e_{act}^{el}\right)\!\!\right];
	\label{eq:v_act_implicit}
\end{equation}
which can be made explicit in the voltage as in \eqref{eq:v_act_explicit}, but is strongly nonlinear and depends on the temperature, both critical factors for solver convergence. 
It can then be replaced at $\lambda = 0$ by its linearized approximation \eqref{eq:linActLoss} computed at a fixed nominal temperature, as suggested in \cite{campanariComparisonFiniteVolume2005}. 

\begin{equation}
	e_{act}^{el} = \frac{RT_{nom}}{\alpha nF}\frac{j}{j_0^{el}}
	\label{eq:linActLoss}
\end{equation}

Regarding the concentration losses, the triple phase boundary pressures are evaluated as in Equation~\eqref{eq:tpbPressure}, calculating for each discretization volume the diffusivity of the mixture's species.
Implementing a model to evaluate diffusivities for any fed mixture introduces several algebraic equations which increase the size of the strong components they might be associated with.
Considering them constant at $\lambda = 0$ together with the partial pressures of the species in the channels is thus beneficial for the initialization procedure.

Regarding the mass and energy balances, the channel model is similar to the one of the heat exchangers and the same explicit state selection is done as in Equations \eqref{eq:dMdt}-\eqref{eq:dUdt}.
The only important difference is that in the fuel cell the flowing streams in the channels are reacting mixtures.
For this reason, the balance equations feature source and sink terms for mass and energy; they are expressed in terms of reaction rates, computed as in Equation~\eqref{eq:reactionRate}:
\begin{equation}
	r_r = k_{0,r}\cdot\exp\qty(-\frac{E_{a,r}}{RT})\cdot p_1^r\cdot p_2^r \cdot \qty(1-\frac{K_{p,r}}{K_{Eq,r}}).
	\label{eq:reactionRate}
\end{equation}
Also in this case, the temperature $T$ is approximated by a fixed $T_{nom}$ at $\lambda = 0$ to decouple the thermal transfer phenomena from the rest of the physical phenomena involved.

On the basis of several numerical experiments using different Modelica tools, we found that the proposed simplifications at $\lambda = 0$ are sufficient to guarantee robust convergence of the resulting smaller nonlinear systems, provided that tearing is applied to them, using only pressures, temperatures, compositions, and currents as iteration variables. This may require to introduce some annotations in the code if the Modelica tool is not able to determine such set of tearing variables automatically.

The fuel cell model originally presented in \cite{depascaliControlOrientedModelica1D2023} was updated to include the previous suggestions;
the full source code of the fuel cell is available on a public GitHub repository \cite{githubSOFC2023} and can be ran in a fully open-source framework with the OpenModelica simulation environment \cite{FritzsonEtAlMIC2020}.

\subsection{The combustor}
The combustor model can be coded as a 0-D volume that processes different incoming flows of fuel, oxidant, and possibly a moderator, to reduce the outlet temperature.
Through the assumption of complete combustion, the flue gases composition is evaluated through simple stoichiometric balances; since unburnt fuel is detrimental for the regular operation of the plant, this assumption is normally fulfilled in reality, by supplying appropriate excess oxygen at the inlet of the component.
Also for this model, the explicit change of state variables for mass and energy balances is performed according to Equations \eqref{eq:dMdt}-\eqref{eq:dUdt}, so the same considerations made for the heat exchanger in Section \ref{sec:hex} still apply.

\subsection{Pressure loss models}
\label{subsec:pDrop}
Pressure losses in pipes are usually described through the Darcy–Weisbach equation; they are characterized by a non-linear quadratic relationship between the mass flow rate and the pressure loss and by a dependency on the temperature through the density of the fluid.
Hence, through the density, fluid dynamics and thermal effects are coupled by this relationship.
As discussed in Section \ref{sec:strategies}, the coupling between these phenomena is detrimental for initialization as it requires the solution of larger strong components and should be avoided at $\lambda = 0$. 
Furthermore, replacing quadratic relationships with linear ones at $\lambda = 0$ removes the need of providing good enough initial guess values to guarantee the solver convergence.

The use of the homotopy method, where actual pressure loss correlation is replaced at $\lambda = 0$ with a linear relationship based on on-design data, such as Equation~\eqref{eq:simplifiedDP}
\begin{equation}
	\Delta p = \frac{\Delta p_{nom}}{w_{nom}}\cdot w
	\label{eq:simplifiedDP}
\end{equation}
can then be a very effective strategy to ease the convergence of the initialization problem. 
Similar considerations apply for other pressure loss components, such as control valves.

As a final remark, thermal power generation systems are designed with the aim of minimizing the pressure losses along the piping and heat exchangers, to avoid efficiency losses. 
Taking this principle in consideration, distributed pressure losses are generally low (a few percentage points of the absolute pressure) and thus they might be modelled employing the proposed linear Equation~\eqref{eq:simplifiedDP} not only as simplified model at $\lambda = 0$, but as the actual friction model, with negligible accuracy losses compared to more accurate models. 
In this case, there is of course no need to employ homotopy-based simplification strategies to ease the initialization problem convergence. 

\section{Overall plant model}
\label{sec:ModelicaPlantInit}
In the previous Section, the components of advanced thermal power generation systems that can be critical for steady-state initialization were discussed, alongside with  strategies employed to facilitate that process.
Once all the components are assembled together to build the plant model, users might still experience convergence issues determined by the size and the complexity of the strong components generated by the components connections.
Moreover, users might ask the software to perform additional tasks during initialization: they may want to obtain a specific full load steady-state output value without knowing a-priori the corresponding inputs to the system or they might want to initialize the plant in off-design conditions. 
The present Section addresses these problems, proposing solutions based on specifically built utility components.

\subsection{The Homotopy Decoupler component}
\label{subsec:homotopyInit}
If convergence failures are experienced when the plant model is assembled, pursuing the divide \& conquer path of decoupling the initialization of different physical sub-systems (e.g. fluid dynamics, thermal, chemical) and of different components at the plant-wide level is of key importance to simulate successfully initialize the system model.

In order to systematically decouple the initialization of each system components in an object-oriented framework, a specific component with inlet and outlet fluid ports is employed. 
This is a general-purpose component that can be used to decouple components of any kind (heat exchangers, turbines, fuel cells, etc.) without the need of modifying their code.

The homotopy decouplers have trivial mass and momentum balance equations: the inlet flow rate is equal to the outlet flow rate and the inlet pressure is equal to the outlet pressure. 
Regarding the outlet specific enthalpy and composition, the idea is to use homotopy again: at $\lambda = 0$, they are fixed to their constant design values, so that the thermal and chemical coupling with the upstream components is removed; at $\lambda = 1$, instead, they are set to be equal to the incoming values at the inlet. 

The use of such components allows to split the original, fully-coupled large nonlinear initialization problem into many smaller problems that can be solved sequentially. 
In fact, if one such homotopy decoupler component is placed at the inlet(s) of each component of the plant (heat exchangers, fuel cells, etc.), the initialization of their thermal and chemical parts may be performed independently. 
Additionally, if the mass and momentum equations of all the involved components are decoupled from the energy and chemical equations at $\lambda = 0$, such thermal and chemical steady-state initialization problems will be set up with known flow rates, which removes a major source of non-linearity in the thermal and chemical equations and thus ensures more robust convergence of the Newton solver.

The main requirement for the use of such blocks is that some (possibly rough or approximated) calculations of the on-design behaviour at each component bounary of the whole system is available a-priori.
 This is normally the case when dealing with dynamic modelling of thermal power generation systems, as the dynamic modelling stage comes after a thermodynamic cycle design phase, in which the values of flow rates, temperatures, compositions and specific enthalpy at each component boundary are known at full load, design conditions.

\subsection{The System Boundary Initialization blocks}
\subsubsection{Requirements for system-wide steady-state initialization}
\label{subseq:requirements}
The dynamic plant model can be used with a simulation tool for multiple purposes: 
\begin{itemize}
	\item it can be employed during the design phase to perform sensibility analysis on specific parameters that might influence the steady-state plant performance; 
	\item it can be used to explore and optimize the off-design operating strategies for the plant;
	\item it may be used to analyze the open-loop dynamic response of the system around on- and off-design steady-state operating points, by running small step response simulations starting from them, or by numerically computing linearized approximations of the system around them;
	\item it may be used to run transient simulations starting from a steady state, possibly including closed-loop control systems to follow the desired setpoints;
	\item it may be exported by the simulation tool as an FMU block \cite{BlochwitzEtAlModelica2011}, to be imported in other simulation tools. Since FMUs embed initialization code derived from the Modelica model, they will be able to initialize themselves to the required on- or off-design steady state independently from the capabilities of the importing tool.
\end{itemize}

In all these cases, suitable on- or off-design steady-state initialization problems must be formulated in the model and solved by the Modelica tool. 
The steady-state initial equations for the individual physical components are specified in the initial equation section of the component models themselves, as seen in the previous Section. 
However, the overall system initialization problem must be completed by adding some extra equations to specify the system boundary values, i.e., which particular steady-state should actually be computed.

An important requirement is that the steady-state initialization code generated by the Modelica tool for each of the previously listed purposes has to be exactly the same. 
This ensures that the successful convergence of the stand-alone steady-state initialization of the plant model guarantees a successful initialization in all those different cases.

As in the case of the homotopy decouplers, a modular solution to this problem should be sought, avoiding the need of altering the code of the physical plant model for each of these different goals.
In this paper, we present reusable general-purpose input and output boundary blocks performing this function, that were designed to initialize the plant in different steady-state conditions to carry out the activities mentioned above in a convenient and effective way.

The physical dynamic model of the plant features input and output connectors corresponding to the actuator and sensor signals. 
This model should be built once and for all, regardless of how it will be used. 
By connecting the system boundary blocks to such input and output connectors, an augmented model is obtained, that can be customized to achieve each of the above-listed purposes by only setting some parameters of those blocks, without touching the physical plant model.

The main goal of the system boundary blocks is thus to provide additional boundary and initial conditions to fully specify a certain steady-state initialization problem. 
However, an additional requirement was considered in this work, concerning input/output normalization. 

Modelica models by convention always employ SI units in first-principles physical equations; this is convenient to guarantee a-priori dimensional consistency of the model equations, but often results in very badly scaled multiple-input, multiple-output (MIMO) models. 
For example, a typical thermal power generation system model may have as inputs and outputs valve flow coefficients in m$^2$ (order of magnitude $10^{-5}-10^{-3}$), thermal and electrical power flows in W (order of magnitude $10^6-10^9$), pressures in Pa (order of magnitude $10^6-10^8$), temperatures in K (order of magnitude $10^3$), chemical compositions (order of magnitude $10^{-3}-1$). 
Thus, the gains of the various input-output pairs in the MIMO model can differ by many orders of magnitude, making its interpretation and further manipulation (e.g. for controller synthesis or model order reduction) unnecessarily difficult.

It is therefore convenient to use the system boundary blocks to also introduce normalization factors, usually set to design values of the corresponding variables, in particular when analyzing the linearized dynamics around an operating point, so that all the inputs and outputs are expressed in per-unit and system dynamic features such as input-output couplings are immediately apparent by looking at the absolute values of the gains.

\subsubsection{Specification of the System Boundary Initialization Blocks}
\label{subseq:specification}
The system boundary blocks have parameters to select one of six possible use scenarios:

\begin{itemize}
	\item \textit{SteadyStateOnDesign} and \textit{SteadyStateOffDesign}: in these configurations, the augmented plant model becomes an autonomous system that can be simulated directly to perform steady-state simulations in on- and off-design conditions. The boundary blocks automatically provide the required input values to the physical plant model, without the need of connecting its inputs to constant signal generators.
	\item \textit{SmallSignalOnDesign} and \textit{SmallSignalOffDesign}: this configuration is meant to analyze the open-loop system response to small variations of the inputs around a certain steady-state operating point, either by simulation of small step responses, or by numerically computing the linearized system dynamics. To support both cases, the system boundary blocks provide the initial equations to specify the initial on- or off-design steady state, as well as input and output connectors that carry normalized \emph{deviations} from that initial steady state. To compute small step responses, the normalized deviation inputs must be connected to small step signal generators and the resulting system should be simulated; otherwise, the augmented system with unconnected top-level inputs and outputs can be directly used by Modelica tools to numerically compute the $A,B,C,D$ matrices of the system, linearized around the specified on- or off-design steady state.
	\item \textit{SimulationOnDesign} and \textit{SimulationOffDesign}: in this configuration, the system boundary blocks provide the initial equations to compute the desired on- or off-design initial steady state, as well as optional normalization of inputs and outputs. The physical plant model is initialized in the specified on- or off-design steady-state conditions \emph{irrespective of the actual initial input values}, thus decoupling the initialization problem of the plant model from the initialization of the rest of the system, e.g., the control system. This is essential to ensure that the structure of the initialization problem in this case is exactly the same as in the previously discusse cases, and also essential to ensure that the model can be exported as an FMU. If one then desires to run simulations that actually start in that specific steady-state, it is necessary to ensure that the initial values of the plant inputs, e.g., as provided by the control system, correspond to the computed steady-state on- or off-design values.
\end{itemize}

Moreover, for each individual system boundary block, the user may choose between two initialization modes:
\begin{itemize}
	\item \textit{Forward initialization}: the \emph{input} boundary blocks fix their steady-state input values to the specified on- or off-design value, while the \emph{output} boundary blocks get the corresponding steady-state values as the results of the computation of the system steady-state.
	\item \textit{Backward initialization}:  the \emph{output} boundary blocks fix their steady-state output values to the specified on- or off-design values, while the \emph{input} boundary blocks get the corresponding steady-state value as the result of the computation of the system steady-state.
\end{itemize}
Note that the initialization problem is just a set of algebraic equations, so there is no inherent causality in it, thus it is conceptually possible to either fix the inputs or the outputs. 
In the base case of a \emph{square} system, with the same number $N$ of inputs and outputs, one can choose to set $N_f$ pairs of input and output boundary blocks in forward mode and $N_b = N-N_f$ pairs of input and output boundary blocks in backward mode. 

Choosing $N_f = N$ for the on-design case means that all the inputs are fixed to the provided design values, and all the outputs are computed as a result. 
However, the dynamic Modelica model of the system can be slightly different from the model used to perform the design calculations, so the computed outputs will typically differ a bit from their design values. 
If one wants to start the simulations or to perform the small-signal analysis around an operating point where the system \emph{outputs} correspond exactly to their design value, $N_b = N$ must be chosen.

Intermediate cases with some input-output pairs in forward mode and others in backward mode can be formulated. 
In particular, this is necessary for those equilibrium points where the linearized $A$ matrix of the system dynamics has eigenvalues in the origin. 
In this case, steady-state input values cannot be chosen arbitrarily but need to fulfill some equilibrium conditions (e.g., the sum of prescribed flow rates into a system must be zero), while the corresponding outputs (e.g. the total mass in the system, or its pressure level) are not fully determined. 
In this case, the appropriate output value(s) (e.g. the system mass or pressure level) should be fixed instead, and the corresponding input values computed by the model, otherwise singular initialization systems would be generated.

If the system is over-actuated, the blocks corresponding to the additional redundant inputs should all be set in forward mode.

When computing off-design steady states, the values of the outputs (e.g. the power output and the controlled temperatures and pressures) are known, as they typically correspond to some known set points, while the exact values of the corresponding inputs are not known. 
In this case, all input-output pairs must be set in backward mode.

Note that different settings of forward/backward initialization modes in general lead to different structures of the corresponding system of initialization equations; in particular, backward initialization may end up in larger strong components at $\lambda = 0$ than forward initialization when the outputs depend on a combination of decoupled physical phenomena. 
For example, if all actuator signals (inputs) are fixed, using the approximations introduced in Sections \ref{sec:components} and \ref{subsec:homotopyInit} allows to first solve the mass balance equations for pressure and flow rates, and then solve the energy balance equations for the temperatures. 
If, instead, the output temperature is fixed, since its value depends on both mass and energy balances, they need to be solved simultaneously. 
Thus, whereas forward initialization problems are very likely to succeed when employing all the strategies discussed in this paper, backward initialization problems are more likely to fail.

This issue can be overcome by first solving the forward problem (ideally only at $\lambda = 0$), which will provide an approximated solution with a high degree of robustness. 
This approximated solution could then be used as an initial guess for the initialization of Newton's method when solving the backward problem, which is likely to converge if the two solutions are not too far apart from each other, regardless of its more complex and nonlinear structure. This can be obtained in Modelica tools by importing the results of the forward calculations to set the initial guesses of the backwards one.

\subsubsection{Implementation}
To describe how the system boundary blocks work, first their variables are presented and then their respective equations are described in Tables \ref{tb:steady_state_table}, \ref{tb:lin_open_table} and \ref{tb:close_table}. 
Recall that parameters in Modelica can either be fixed, in which case their values are known at the beginning of the initialization process, or unknown, in which case they are included in the set of unknowns of the initialization process and will be computed by it.

The code of the \emph{input} block features the following variables and parameters:
\begin{itemize}
	\item $u_{in}$: connector input variable fed to the initialization block by an external signal source;
	\item $u_{out}$: connector output variable that serves as an input for the connected component of the plant, normally an actuator or some disturbance generator;
	\item $u_{norm}$: fixed parameter for (optional) de-normalization of $u_{in}$;
	\item $u_{des}$: fixed parameter representing the \emph{a-priori} known on-design value of the variable $u_{out}$;
	\item $u_{des,calc}$: unknown parameter representing the \emph{calculated} on-design value of $u_{out}$;
	\item $u_{offdes,calc}$: unknown parameter representing the \emph{calculated} off-design value of $u_{out}$;
\end{itemize}

The code of the \emph{output} block features the following variables and parameters:
\begin{itemize}
	\item $y_{in}$: connector input variable received by the initialization block from the connected component of the plant, usually a sensor;
	\item $y_{out}$: connector output variable of the initialization block;
	\item $y_{norm}$: fixed parameter for (optional) normalization of $y_{out}$;
	\item $y_{des}$: fixed parameter representing the \emph{desired} on-design value of the variable $y_{in}$;
	\item $y_{offdes}$: fixed parameter representing the \emph{desired} off-design value of the variable $y_{in}$;
	\item $y_{des,calc}$: unknown parameter representing the \emph{calculated} on-design value of the variable $y_{in}$;
	\item $y_{offdes,calc}$: unknown parameter representing the \emph{calculated} off-design value of the variable $y_{in}$;
\end{itemize}

\begin{table}[h]
	\centering
	\caption{Table representing the equations of the input-output blocks for the \textit{SteadyStateOnDesign} and \textit{SteadyStateOffDesign} scenarios}
	\begin{tabular}{ccc|c|c|}
		\cline{4-5}
		&                                        &        & EQUATIONS & \begin{tabular}[c]{@{}c@{}}INITIAL\\ EQUATIONS\end{tabular} \\ \hline
		\multicolumn{1}{|c|}{\multirow{2}{*}{FWD}}  & \multicolumn{1}{c|}{\multirow{2}{*}{ON}}  & IN  & $u_{out} = u_{des,calc}$        & $ \begin{aligned} & u_{des,calc} = u_{des} \\ & u_{offdes,calc} = u_{des} \end{aligned} $                                                          \\ \cline{3-5} 
		\multicolumn{1}{|c|}{}                          & \multicolumn{1}{c|}{}                            & OUT & $y_{out} = y_{in}$        & $\begin{aligned}&y_{des,calc} = y_{des}\\&y_{offdes,calc} = y_{des}\end{aligned}$                                                          \\ \hline
		\multicolumn{1}{|c|}{\multirow{4}{*}{BWD}} & \multicolumn{1}{c|}{\multirow{2}{*}{ON}}  & IN  & $u_{out} = u_{des,calc}$       & $u_{offdes,calc} =  u_{des}$                                                          \\ \cline{3-5} 
		\multicolumn{1}{|c|}{}                          & \multicolumn{1}{c|}{}                            & OUT & $y_{out} = y_{in}$        & $\begin{aligned} &y_{in} = y_{des}\\	&y_{des,calc} = y_{des} \\
			&y_{offdes,calc} = y_{des} \end{aligned}$                                                          \\ \cline{2-5} 
		\multicolumn{1}{|c|}{}                          & \multicolumn{1}{c|}{\multirow{2}{*}{OFF}} & IN  & $u_{out} = u_{offdes,calc}$        & $u_{des,calc} = u_{des} $                                                         \\ \cline{3-5} 
		\multicolumn{1}{|c|}{}                          & \multicolumn{1}{c|}{}                            & OUT & $y_{out} = y_{in}$        & $\begin{aligned} &y_{in} = \text{homotopy(}y_{offdes}, y_{des}\text{)}\\
			&y_{des,calc} = y_{des}\\
			&y_{offdes,calc} = y_{offdes} \end{aligned}$                                                          \\ \hline
	\end{tabular}
	\label{tb:steady_state_table}
\end{table}

Table \ref{tb:steady_state_table} contains the equations (active during both initialization and simulation) and initial equations (active only during initialization) that characterize the input (IN) and output (OUT) blocks when they are employed in \emph{SteadyStateOnDesign} and \emph{SteadyStateOffDesign} scenarios, in \emph{forward} and \emph{backward} modes.
For a forward (FWD) on-design (ON) initialization, the initial equations in the IN block set the auxiliary parameters $u_{des,calc}$ and $u_{offdes,calc}$ to be equal to the design value $u_{des}$. 
During simulation, $u_{out}$ is set equal to the computed $u_{des,calc}$ and supplies the constant input signal to the connected component (actuator). The input $u_{in}$ does not play a role in this mode and can actually be removed from the model by using conditional connectors.
Regarding the OUT block, the calculated on- and off-design values are again set to the design value in the initial equations, even though they actually do not play any role in the equations, whereas, during simulation, the block output $y_{out}$ is equal to the block input $y_{in}$, i.e., the connected plant sensor output, which will be computed as a result of the initialization system.

For a backward (BWD) on-design (ON) initialization, the equations of the blocks are the same as for the forward mode, while the \emph{initial} equations change. 
In particular, since the design condition is now fixed at the output, the initial equation $u_{des,calc} = u_{des}$ is removed from the IN block, while the initial equation $y_{in} = y_{des}$ is added to the OUT block. 
When considering the overall initialization problem, $u_{out}$ (i.e. the actuator signal value) will be computed by solving the steady-state plant equations backward with fixed output, and assigned to $u_{des,calc}$. 
As long as BWD mode is set on pair of input/output boundary blocks, the overall number of initial equations remains the same, so the initialization problem remains balanced.

The backward (BWD) off-design initialization (OFF) is similar to the BWD ON case, with two differences. 
The first is that the equation in the IN block now sets the block output $u_{out}$ to the calculated off-design value $u_{offdes,calc}$ instead of the calculated on-design value $u_{des,calc}$. 
The second is that the OUT block input $y_{in}$, i.e. the plant sensor output, is not directly set to $y_{offdes}$ but gradually moved from the on-design to the off-design value via homotopy. 
The motivation of this homotopy transformation is that all the simplifications introduced at $\lambda = 0$ in Sections \ref{sec:components} and \ref{subsec:homotopyInit} are valid \emph{in on-design conditions}, hence also the backward problem should first be solved in those conditions; alongside the homotopy transformation, the solution of the problem is gradually brought to the desired off-design conditions. 
This strategy completely avoids the need to provide guess values for off-design initialization, relying exclusively upon on-design system data.

It should be noticed that the transition of the imposed outputs to off-design conditions and the one from the simplified to the actual form of the model equations are carried out simultaneously along the homotopy transformation, which entails additional nonlinearity that may required shorter $\lambda$ steps to successfully achieve convergence. 
In principle, one could perform the two homotopy transformations sequentially: first transform the simplified model into the actual one \emph{in on-design conditions}, then move the conditions to the off-design ones. 
In practice, experience has shown that this is apparently not necessary to achieve successful convergence.

\begin{table}[tbph]
	\centering
	\caption{Table representing the equations of the input-output blocks for the \textit{SmallSignalOnDesign} and \textit{SmallSignalOffDesign} scenarios}
	\begin{tabular}{ccc|c|c|}
		\cline{4-5}
		&                                        &        & EQUATIONS & \begin{tabular}[c]{@{}c@{}}INITIAL\\ EQUATIONS\end{tabular} \\ \hline
		\multicolumn{1}{|c|}{\multirow{2}{*}{FWD}}  & \multicolumn{1}{c|}{\multirow{2}{*}{ON}}  & IN  & $u_{in} = \frac{u_{out} - u_{des,calc}}{u_{norm}}$        & $ \begin{aligned} & u_{des,calc} = u_{des} \\ & u_{offdes,calc} = u_{des} \end{aligned} $                                                          \\ \cline{3-5} 
		\multicolumn{1}{|c|}{}                          & \multicolumn{1}{c|}{}                            & OUT & $y_{out} = \frac{y_{in} - y_{des,calc}}{y_{norm}}$        & $\begin{aligned}&y_{des,calc} = y_{des}\\&y_{offdes,calc} = y_{des}\end{aligned}$                                                          \\ \hline
		\multicolumn{1}{|c|}{\multirow{4}{*}{BWD}} & \multicolumn{1}{c|}{\multirow{2}{*}{ON}}  & IN  & $u_{in} = \frac{u_{out} - u_{des,calc}}{u_{norm}}$       & $u_{offdes,calc} =  u_{des}$                                                          \\ \cline{3-5} 
		\multicolumn{1}{|c|}{}                          & \multicolumn{1}{c|}{}                            & OUT & $y_{out} = \frac{y_{in} - y_{des,calc}}{y_{norm}}$        & $\begin{aligned} &y_{in} = y_{des}\\	&y_{des,calc} = y_{des} \\
			&y_{offdes,calc} = y_{des} \end{aligned}$                                                          \\ \cline{2-5} 
		\multicolumn{1}{|c|}{}                          & \multicolumn{1}{c|}{\multirow{2}{*}{OFF}} & IN  & $u_{in} = \frac{u_{out} - u_{offdes,calc}}{u_{norm}}$        & $u_{des,calc} = u_{des} $                                                         \\ \cline{3-5} 
		\multicolumn{1}{|c|}{}                          & \multicolumn{1}{c|}{}                            & OUT & $y_{out} = \frac{y_{in} - y_{offdes,calc}}{y_{norm}}$        & $\begin{aligned} &y_{in} = \text{homotopy(}y_{offdes}, y_{des}\text{)}\\
			&y_{des,calc} = y_{des}\\
			&y_{offdes,calc} = y_{offdes} \end{aligned}$                                                          \\ \hline
	\end{tabular}
	\label{tb:lin_open_table}
\end{table}
\begin{table}[t!]
	\centering
	\caption{Table representing the equations of the input-output blocks for the \textit{SimulationOnDesign} and \textit{SimulationOffDesign} scenarios}
	\begin{tabular}{ccc|c|c|}
		\cline{4-5}
		&                                        &        & EQUATIONS & \begin{tabular}[c]{@{}c@{}}INITIAL\\ EQUATIONS\end{tabular} \\ \hline
		\multicolumn{1}{|c|}{\multirow{2}{*}{FWD}}  & \multicolumn{1}{c|}{\multirow{2}{*}{ON}}  & IN  & $\begin{aligned}u_{out} &= \\&\text{if initial() then }\\&\;\;u_{des,calc}\\& \text{else }\\ &\;\;u_{in}\end{aligned}$        & $ \begin{aligned} & u_{des,calc} = u_{des} \\ & u_{offdes,calc} = u_{des} \end{aligned} $                                                          \\ \cline{3-5} 
		\multicolumn{1}{|c|}{}                          & \multicolumn{1}{c|}{}                            & OUT & $y_{out} = y_{in}$        & $\begin{aligned}&y_{des,calc} = y_{des}\\&y_{offdes,calc} = y_{des}\end{aligned}$                                                          \\ \hline
		\multicolumn{1}{|c|}{\multirow{4}{*}{BWD}} & \multicolumn{1}{c|}{\multirow{2}{*}{ON}}  & IN  & $\begin{aligned}u_{out} &= \\&\text{if initial() then }\\&\;\;u_{des,calc}\\& \text{else }\\ &\;\;u_{in}\end{aligned}$       & $u_{offdes,calc} =  u_{des}$                                                          \\ \cline{3-5} 
		\multicolumn{1}{|c|}{}                          & \multicolumn{1}{c|}{}                            & OUT & $y_{out} = y_{in}$        & $\begin{aligned} &y_{in} = y_{des}\\	&y_{des,calc} = y_{des} \\
			&y_{offdes,calc} = y_{des} \end{aligned}$                                                          \\ \cline{2-5} 
		\multicolumn{1}{|c|}{}                          & \multicolumn{1}{c|}{\multirow{2}{*}{OFF}} & IN  & $\begin{aligned}u_{out} &= \\&\text{if initial() then }\\&\;\;u_{offdes,calc}\\& \text{else }\\ &\;\;u_{in}\end{aligned}$        & $u_{des,calc} = u_{des} $                                                         \\ \cline{3-5} 
		\multicolumn{1}{|c|}{}                          & \multicolumn{1}{c|}{}                            & OUT & $y_{out} = y_{in}$        & $\begin{aligned} &y_{in} = \text{homotopy(}y_{offdes}, y_{des}\text{)}\\
			&y_{des,calc} = y_{des}\\
			&y_{offdes,calc} = y_{offdes} \end{aligned}$                                                          \\ \hline
	\end{tabular}
	\label{tb:close_table}
\end{table}

Table~\ref{tb:lin_open_table} shows the equations for \emph{SmallSignalOnDesign} and \emph{SmallSignalOffDesign} cases.
The initial equations are the same as in the case of the steady-state simulation. 
The equations are modified to define the IN block input signal and the OUT block output signal as normalized deviations with respect to the calculated on- or off-design value.

Table \ref{tb:close_table} shows the equations for the \textit{SimulationOnDesign} and \textit{SimulationOffDesign} modes assuming input/output normalization is not enabled; the details of that option are trivial and omitted for the sake of brevity. 
The \emph{initial equations} are the same as the ones employed for the previously discussed cases. 
The \emph{equations} of the IN blocks instead are changed to ensure that, during initialization, the block input is ignored and the output is the calculated on- or off-design value, while it is passed unchanged during simulation.

A small Modelica library containing the implementation of the system boundary initialization blocks is available as open-source on GitHub \cite{boundInitBlock}.
The functionality of the system boundary initialization blocks was successfully tested in all the different scenarios and modes presented in Section~\ref{subseq:specification}, using very simple process model for the demonstration.
The test cases are not discussed in this paper for the sake of brevity; the interested reader is encouraged to download and install the OpenModelica tool, then install the \emph{BoundaryInitBlocks} library using the package manager and then explore and run them.

\section{Industrial test case: the SOS-CO$_2$ cycle}
\label{sec:sosco2}
The concepts described in the previous three Sections were applied to build and initialize the dynamic model of the Solid Oxide Semi-Closed CO\tsub{2} (SOS-CO$_2$) cycle \cite{scaccabarozziSolidOxideSemiclosed2021}, a technology that belongs to the family of the oxy-combustion power plants.
SOS-CO$_2$ combines a fuel cell with a recuperated oxy-combustion Brayton cycle and may reach an outstanding efficiency of 75.7\%, including carbon capture, in the most extreme configuration. 
The test case considered here is a dynamic model of the low-risk configuration defined in \cite{scaccabarozziSolidOxideSemiclosed2021} and detailed in \cite{depascali2024}, featuring a maximum pressure of about \SI{30.0}{\bar} and a minimum pressure slightly above the atmospheric one. 
Note that the information available in \cite{scaccabarozziSolidOxideSemiclosed2021} only concerned on-design operating data for the system, so one of the goals of the developed model was also to compute off-design steady-state operating point, allowing to study suitable off-design operating strategies.

\begin{figure}[tbph!]
	\includegraphics[width=\linewidth]{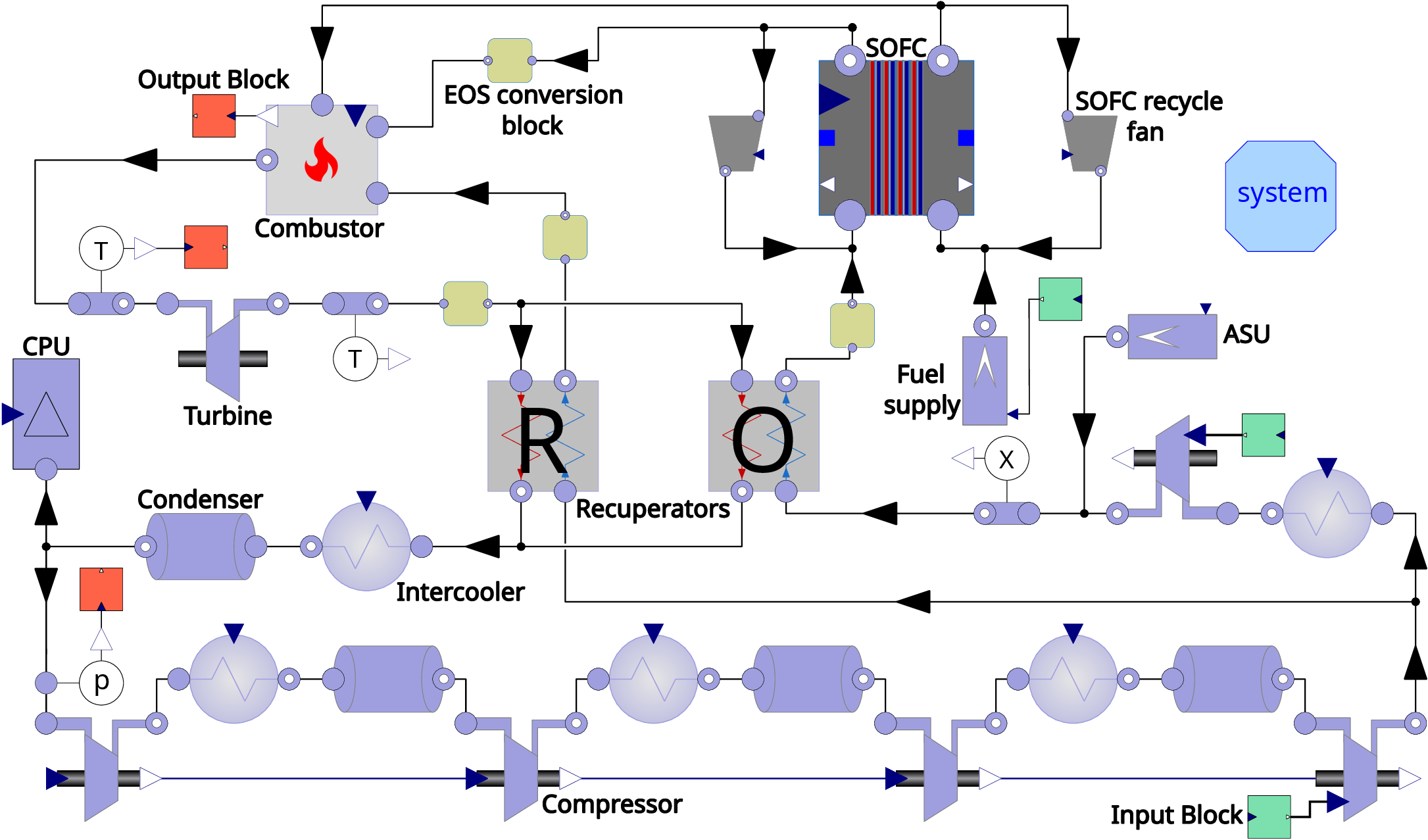}
	\caption{Diagram of the SOS-CO\tsub{2} cycle in Modelica with initialization blocks pairs connected to some of the input (green blocks) and outputs (orange blocks) of the system.} 
	\label{fig:modelicaPlantSchemeBlocks}
\end{figure}

The cycle is composed by the following components and its flowsheet is shown in Figure \ref{fig:modelicaPlantSchemeBlocks}:
\begin{itemize}
	\item{a turbine receiving the exhaust gases from the oxy-combustor to generate electrical power through flow expansion;}
	\item{recuperators to recover the thermal power from the hot flue gases and preheat the moderator and oxidant flows;}
	\item{a compression and purification unit (CPU) to extract from the cycle the CO\tsub{2} produced by the oxidation of the fuel;}
	\item{a train of centrifugal compressors with intercoolers and condensers to bring the moderator stream to the maximum pressure of the cycle;}
	\item{a centrifugal compressor and an Air Separation Unit (ASU) to mix part of the moderator flow with a pure oxygen stream to get the oxidant stream;}
	\item{a Solid Oxide Fuel Cell (SOFC) with recycle fans to convert the natural gas fuel supply to hydrogen, water vapour and carbon dioxide, and to partially oxidize it with the oxidant flow through electro-chemical reactions;}
	\item{an oxy-combustor receiving the anodic exhaust flow and the oxidant from the SOFC, and the moderator flow to completely burn the remaining hydrogen, producing a mixture of carbon dioxide and water vapour.}
\end{itemize}

The overall plant model accounts for over 110000 differential-algebraic equations; the size and the complexity of the nonlinear systems to be solved for steady state initialization proved to be such that the adoption of \emph{all} the strategies presented in the paper was necessary to achieve robust and efficient steady-state initialization. 
The nonlinear behaviour of the fuel cell and of the recuperators proved to be particularly challenging, if not tamed by the presented simplification strategies.

In particular, the homotopy decoupler blocks introduced in Section~\ref{sec:ModelicaPlantInit} turned out to be essential to decouple the components' initialization and to explore off-design operating points.

The SOS-CO$_2$ plant model was successfully initialized in on-design conditions using both Dymola and OpenModelica simulation software; at the time of this writing, off-design initializations were performed successfully only using Dymola. 
This allowed to explore multiple partial load control policies, e.g. switching to turbine inlet temperature control to turbine outlet temperature control below a certain operating load, as well as suitably partitioning the power generation between the fuel cell and the turbine.

Moreover, the small-signal open-loop behaviour of the model around different on- and off-design operating points was successfully studied by running small step responses and by directly computing linearized models; the latter ones have a very high order (more than 1500 state variables), so they were suitably simplified using standard linear model-order reduction techniques such as balance reduction. 
Eventually, based on this information, a plant-wide control system was designed and successfully simulated in realistic scenarios, which all start from steady-state conditions. 
Unfortunately, the details of such analysis are confidential and cannot be presented here. 
It is anyway possible to state that the use of the presented strategies for initialization was a crucial factor in the success of that activity.

\section{Conclusions}
\label{sec:conclusions}
The forthcoming energy transition calls for a new generation of thermal power generation systems with low or zero emission and highly flexible operation. 
Dynamic modelling and simulation is a key enabling factor in this field, as controlling such plants is a difficult task for which there is no previous experience and very short design times are expected. 
The steady-state initialization of those dynamic models is an essential step in the design process, but is unfortunately a difficult task which involves the numerical solution of large systems of nonlinear equations with iterative Newton methods, which is often prone to numerical failures.

In this work, several strategies and methodologies to successfully achieve the steady-state initialization of dynamic models of advanced thermal power generation systems were presented.

The main tool proposed in this paper to address this problem is the use of homotopy-based initialization strategies, that allow to solve a simplified problem first, and then reach the solution of the actual initialization problem gradually, using a homotopy transformation. 
The two main principles guiding the formulation of the simplified models are: a) making the problems less nonlinear, so that the convergence of Newton's method is less dependent (or not dependent at all) from an accurate set-up of the initial guess values for the unknowns; b) decoupling the problem into many smaller problem to be solved sequentially, which are ususally faster and more robust to solve than the original coupled one.

Strategies to achieve these goals at the component level and at the system level were discussed. 
At the component level, this requires writing suitably simplified equations, which were discussed with reference to typical thermal power generation system components. 
At the system level, general purpose components to decouple the overall system initialization into a number of smaller problems were discussed. 
General-purpose system boundary blocks were then introduced, which allow to formulate a wide range of different initialization problems in a modular way for a given physical model of the plant, and to solve them reliably.

Finally, the successful application of such techniques to a challenging use case, the dynamic model of the SOS-CO$_2$ case, was briefly presented.

The strategies discussed in this paper are applicable to a wide range of equation-based, object-oriented dynamic models of thermal power generation systems. They are demonstrated in the case of the stand-alone fuel cell in an open-source package available on GitHub  \cite{githubSOFC2023}, which can be ran with the open-source OpenModelica simulation environment \cite{FritzsonEtAlMIC2020}.
The system boundary blocks have a very general applicability and are also available as open-source on GitHub \cite{boundInitBlock}.

\section*{CRediT authorship contribution statement}
\textbf{Matteo Luigi De Pascali}: Methodology, Software, Validation, Investigation, Writing Original Draft. \textbf{Francesco Casella}: Conceptualization, Methodology, Writing Original Draft, Writing Review and Editing, Supervision, Funding Acquisition

\section*{Declaration of Competing Interest}
The authors declare that they have no known competing financial interests or personal relationships that could have appeared to influence the work reported in this paper.

\section*{Acknowledgements}
This work was supported by the Italian Ministry of University and Research under the PONRI program (Programma Operativo Nazionale "Ricerca e Innovazione" 2014-2020) and by ENI S.p.A. through the Joint Research Center with Politecnico di Milano. 

The authors are grateful to Dylan Vandoni of ENI for providing the industrial test case and to Emanuele Martelli and Matteo Martinelli of the Department of Energy of Politecnico di Milano for providing the steady-state design data of the SOS-CO$_2$ plant and for their continous support throughout the development of the dynamic model of that plant. The authors are also grateful to former master's students Marcelo Muro Alvarado and Giovanni Mangola who worked on early versions of the boundary initialization concept.




\bibliographystyle{elsarticle-num}
\bibliography{biblio.bib,oom.bib}
%
%
%
\end{document}